\documentclass[a4paper,11pt]{article}
\pdfoutput=1 
\usepackage{jheppub} 
\usepackage[T1]{fontenc} 
\usepackage{caption}
\usepackage{subcaption}
\usepackage{hyperref}
\usepackage{mwe}
\title{\boldmath One-loop corrections to the celestial chiral algebra from Koszul Duality}
 \author{Víctor E. Fernández}
 \affiliation{Department of Physics, University of Washington, Seattle, WA, USA}
\emailAdd{vfer@uw.edu}
\abstract{We consider self-dual Yang-Mills theory (SDYM) in four dimensions and its lift to holomorphic BF theory on twistor space. Following the work of Costello and Paquette, we couple SDYM to a quartic axion field, which guarantees associativity of the (extended) celestial chiral algebra at the quantum level. We demonstrate how to reproduce their one-loop quantum deformation to the chiral algebra using Koszul duality.}

\begin{document} 
\maketitle
\flushbottom

\section{Introduction}
Collinear singularities in the scattering amplitudes of gauge theory have been suggested to be encoded in a two-dimensional CFT by the celestial holography program (see \cite{LECTURES_ON_THE_INFRARED_STRUCTURE_OF_GRAVITY_AND_GAUGE_THEORY, LECTURES_ON_CELESTIAL_HOLOGRAPHY, LECTURES_ON_CELESTIAL_AMPLITUDES} and references therein). In the case of self-dual gauge theory with only states of positive helicity, this was shown to be true at tree level \cite{HOLOGRAPHIC_SYMMETRY_ALGEBRAS_FOR_GAUGE_THEORY_AND_GRAVITY} and later at one-loop level \cite{PERTURBATIVELY_EXACT_ASYMPTOTIC_SYMMETRY_OF_QUANTUM_SELF_DUAL_GRAVITY}. In \cite{ON_THE_ASSOCIATIVITY_OF_ONE_LOOP_CORRECTIONS_TO_THE_CELESTIAL_OPE}, it was shown that the one-loop collinear singularities \cite{Bern_1994, Kosower_1999, Bern_2005} of the self-dual limit of pure gauge theory with states of both helicities did not lead to a consistent chiral algebra, as associativity did not persist past tree-level. This was a consequence of the twistor theory uplift of self-dual Yang Mills (SDYM) suffering from a gauge anomaly associated to a box diagram. For certain gauge groups, this can be remedied by a Green-Schwarz mechanism with the introduction of a quartic axion field \cite{QUANTIZING_HOLOMORPHIC_FIELD_THEORIES_ON_TWISTOR_SPACE}. The one-loop deformation in the axion-coupled theory satisfies associativity and thus preserves the structure of the universal collinear singularitites as a chiral algebra at one-loop level. We refer to this as the (extended) celestial chiral algebra. \\ \\
The 1-loop corrections to the chiral algebra were calculated in \cite{ON_THE_ASSOCIATIVITY_OF_ONE_LOOP_CORRECTIONS_TO_THE_CELESTIAL_OPE} using known 4d collinear splitting amplitudes and the requirement of associativity. In this paper, we will use Koszul duality (see \cite{KOSZUL_DUALITY_IN_QFT} for a review) to obtain these 1-loop corrections. Koszul duality is a mathematical notion which in essence takes an algebra $A$ satisfying certain conditions and produces a new Koszul dual algebra $A^{!}$ such that $(A^{!})^{!} = A$. In \cite{TWISTED_SUPERGRAVITY_AND_KOSZUL_DUALITY}, it was shown that we can extend the framework of Koszul duality to act on chiral algebras. For recent mathematical developments, see \cite{QUADRATIC_DUALITY_FOR_CHIRAL_ALGEBRAS} and references therein. In this case, given a chiral algebra, Koszul duality produces a new Koszul dual chiral algebra.  \\ \\
Suppose we have some twist of a supersymmetric quantum field theory (see \cite{TASI_LECTURES_ON_THE_MATHEMATICS_OF_STRING_DUALITIES} for a review) on $\mathbb{C} \times (\mathbb{C})^n$ which is BRST-invariant, holomorphic along $\mathbb{C}$ and, in general, a combination of holomorphic and topological in the other directions. Let $A$ be the differential-graded (DG) chiral algebra of (not necessarily gauge-invariant) local operators restricted to $\mathbb{C}$, with the BRST operator as its differential and a known OPE. The Koszul dual $A^{!}$ is then defined as the universal chiral algebra that can be coupled to our theory as the algebra of operators of a holomorphic defect wrapping $\mathbb{C}$. We demand that this coupling be BRST invariant, which results in contraints on the OPEs between the operators of $A^{!}$. This procedure can be modeled by Feynman diagrams. This interpretation allows us to work in perturbation theory, ensuring BRST invariance order-by-order.
\\ \\
The structure of this paper is as follows:  
\begin{itemize}
  \item \textbf{Section 2}. We review 6d holomorphic BF theory in the BV formalism and its BRST variations. 
  \item \textbf{Section 3 and 4}. Following \cite{TWISTED_SUPERGRAVITY_AND_KOSZUL_DUALITY}, we briefly review the celestial chiral algebra and its tree-level OPEs from the point of view of Koszul duality.\footnote{Note that at tree-level, we can neglect the fact that 6d holomorphic BF theory fails to be BRST-invariant at 1-loop.}
  \item \textbf{Section 5 and 6}. We review the inclusion of the quartic axion field to the twistorial theory, its BV action functional and BRST variations following \cite{CELESTIAL_HOLOGRAPHY_MEETS_TWISTED_HOLOGRAPHY_4D_AMPLITUDES_FROM_CHIRAL_CORRELATORS}. We also discuss the introduction of the corresponding two towers of generators to the (extended) celestial chiral algebra. From the fields of this theory, one can construct "bulk" local operators, the restriction of which to the twistor $\mathbb{CP}^1$ will give us $A$.
  \item \textbf{Section 7}. We calculate the 1-loop corrections to the (extended) celestial chiral algebra using Koszul duality. Note that a similar computation in the context of self-dual gravity coupled to a $4^{\text{th}}$-order gravitational axion was also recently performed in \cite{ON_THE_ASSOCIATIVITY_OF_1_LOOP_CORRECTIONS_TO_THE_CELESTIAL_OPERATOR_PRODUCT_IN_GRAVITY}. We relegate details of some holomorphic integrals to the Appendices. 
\end{itemize}
\section{Holomorphic BF Theory}
The holomorphic BF-type action on twistor space is given by\footnote{We use a different normalization convention to that of \cite{ON_THE_ASSOCIATIVITY_OF_ONE_LOOP_CORRECTIONS_TO_THE_CELESTIAL_OPE}. In particular, we normalize our integrals by a factor of $\frac{1}{2 \pi i}$.  \label{footnote_1}}
\begin{equation}
   S[\mathcal{A}, \mathcal{B}] = \bigg(\frac{1}{2 \pi i}\bigg) \underset{\mathbb{PT}}{\int} \text{Tr}(\mathcal{B} F^{0,2}(\mathcal{A})) = \bigg(\frac{1}{2 \pi i}\bigg)\underset{\mathbb{PT}}{\int} \text{Tr}(\mathcal{B} \overline{\partial} \mathcal{A}+\frac{1}{2} \mathcal{B} [\mathcal{A}, \mathcal{A}])
\end{equation}
where the field content of the theory is $\mathcal{A} \in \Omega^{0,1}(\mathbb{PT}, \mathfrak{g})$ and $\mathcal{B} \in \Omega^{3,1}(\mathbb{PT}, \mathfrak{g})$ for $\mathfrak{g}$ a complex semi-simple Lie algebra. The fields $\mathcal{A}$ and $\mathcal{B}$ are subject to two gauge variations with generators $\chi \in \Omega^{0,0}(\mathbb{PT}, \mathfrak{g})$ and $\nu \in \Omega^{3,0}(\mathbb{PT}, \mathfrak{g})$
\begin{equation}
    \delta \mathcal{A} = \overline{\partial} \chi + [\mathcal{A}, \chi] \quad \quad \delta \mathcal{B} = \overline{\partial} \nu + [\mathcal{B}, \chi].
\end{equation}
In the BV formalism, we extend  $\mathcal{A}$ to a field in $\Omega^{0,*}(\mathbb{PT}, \mathfrak{g})[1]$ and $\mathcal{B}$ to one in  $\Omega^{3,*}(\mathbb{PT}, \mathfrak{g})[1]$, where [1] denotes a shift in ghost number so that fields in Dolbeault degree j are in ghost number $1-j$. In particular, the $(*,1)$ component of these polyform fields correspond to the physical fields. Explicitly, the resulting polyform fields are written as 
\begin{equation}
    \tilde{\mathcal{A}} = \chi + \mathcal{A} + \mathcal{B}^{\vee} + \nu^{\vee} \quad \quad \Tilde{\mathcal{B}} = \nu + \mathcal{B} + \mathcal{A}^{\vee}+\chi^{\vee}
\end{equation}
where $( \cdot )^{\vee}$ denotes the antifield of $( \cdot )$. The resulting BV action is 
\begin{equation}
    S[\tilde{\mathcal{A}},\tilde{\mathcal{B}}] = \bigg(\frac{1}{2 \pi i}\bigg)\underset{\mathbb{PT}}{\int} \text{Tr}(\tilde{\mathcal{B}} \overline{\partial} \tilde{\mathcal{A}}+\frac{1}{2} \tilde{\mathcal{B}} [\tilde{\mathcal{A}}, \tilde{\mathcal{A}}]).
\end{equation}
Writing the action out in terms of the components of the polyform fields, the holomorphic BF action obtains the additional terms:
\begin{equation}
     \bigg(\frac{1}{2 \pi i}\bigg)\underset{\mathbb{PT}}{\int} \text{Tr} \bigg(
     \mathcal{A}^{\vee}(\overline{\partial} \chi + [\mathcal{A}, \chi]) +\mathcal{B}^{\vee} (\overline{\partial} \nu + [\mathcal{A}, \nu] + [\mathcal{B}, \chi]) +\frac{1}{2} \chi^{\vee} [ \chi, \chi]+ \nu^{\vee} [\chi, \nu]  \bigg).
\end{equation}
The BRST transformation of the fields are encoded in their equations of motion. This can then be easily read off from this form of the action, as $\delta_{( \cdot )^{\vee}}$ yields the equation of motion for $( \cdot )$:
\begin{equation}
        \delta \mathcal{A} = \overline{\partial} \chi + [\mathcal{A}, \chi] \quad \quad \delta \mathcal{B} = \overline{\partial} \nu + [\mathcal{A}, \nu] + [\mathcal{B},\chi] \quad \quad 
        \delta \chi = \frac{1}{2} [\chi, \chi] \quad \quad \delta \nu = [\chi, \nu].
\end{equation}
\\ At the classical level, the holomorphic BF action reduces to 4d SDYM on $\mathbb{R}^4$:
\begin{equation}
    \underset{\mathbb{R}^4}{\int} \text{Tr}(BF(A)_{-})
\end{equation}
where the field content is $A \in \Omega^{1}(\mathbb{R}^4,\mathfrak{g})$ and  $B \in \Omega_{-}^{2}(\mathbb{R}^4,\mathfrak{g})$ \cite{ON_SELF_DUAL_GAUGE_FIELDS}. The reduction from twistor theory incorporates states of both helicities \cite{TWISTOR_ACTIONS_FOR_NON_SELF_DUAL_FIELDS}. 
 \section{Chiral Algebra}
 \begin{figure}[t]
    \centering
    \begin{subfigure}[b]{0.3\textwidth}
    \centering
    \includegraphics[scale=0.3]{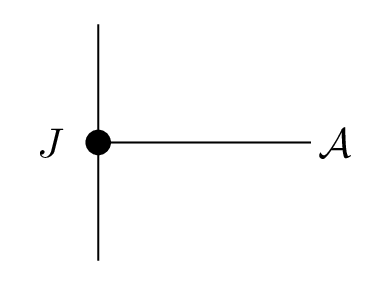}
    \end{subfigure}
    \begin{subfigure}[b]{0.3\textwidth}
\centering
         \includegraphics[scale=0.3]{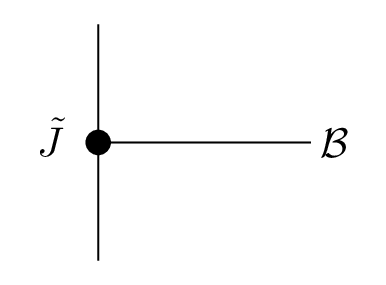}
    \end{subfigure}
     \caption{}
     \label{figure_1}
\end{figure}
Consider a defect along a holomorphic plane $\mathbb{C} \subset \mathbb{C}^3$ and choose coordinates $z$, $v^1$, $v^2$ so that the plane is located at $v^i = 0$. Since anomalies are local, working with this local model for $\mathbb{CP}^1\subset \mathbb{PT}$ is sufficient. The most general way that this defect theory couples to holomorphic BF theory is 
\begin{equation}
    \sum_{m,n \geq 0} \bigg( \frac{1}{2 \pi i} \bigg) \frac{1}{m! n!} \underset{\mathbb{C}}{\int} d^2 z \bigg(J_a[m,n] \partial_{v^1}^m \partial_{v^2}^n \mathcal{A}_{\overline{z}}^a + \tilde{J}_a[m,n] \partial_{v^1}^m \partial_{v^2}^n \mathcal{B}_{\overline{z}}^a
    \bigg)
\end{equation}
in terms of some general defect operators $J$ and $\tilde{J}$. We will assume the defect theory whose algebra of operators $B$ includes $J$ and $\tilde{J}$ is not itself a 2d gauge theory, i.e. $B$ is an ordinary algebra, not a DGA. This coupling diagramatically takes the form of Figure 1. The $J$ and $\tilde{J}$ towers of states form the celestial chiral algebra for SDYM with states of both helicities. Their scaling dimension is $-(m+n)$ and $-2-(m+n)$, and their spin is $1-\frac{m+n}{2}$ and $-1-\frac{m+n}{2}$, respectively.\footnote{Dimension here corresponds to the charge of the operator under scaling of $\mathbb{R}^4$. Spin refers to holomorphic 2d conformal weight.}  
\\ \\
We demand that this coupling be gauge invariant. This requires all anomalous (i.e., BRST-non-invariant) Feynman diagram contributions to cancel order-by-order in perturbation theory. This will lead to constraints for the OPEs between the defect operators. For more details, see section 6 of \cite{TWISTED_SUPERGRAVITY_AND_KOSZUL_DUALITY} and the appendix of \cite{ASPECTS_OF_OMEGA_DEFORMED_M_THEORY}.
\section{Tree-Level OPEs}
The tree-level Feynman diagrams for the bulk-defect interactions are illustrated in Figure 2. The contribution of the left-most diagram corresponds to the gauge variation of the $J_a \mathcal{A}^a_{\overline{z}}$ coupling
\begin{equation}
    \sum_{l,k \geq 0} \bigg( \frac{1}{2 \pi i} \bigg) \frac{1}{l! k!}\underset{\mathbb{C}}{\int} d^2 z f^a_{bc} J_a[l,k] \partial_{v^1}^l \partial_{v^2}^k (\chi^b \mathcal{A}_{\overline{z}}^c),
\end{equation}
where we've integrated by parts and used $\partial_{\overline{z}}J_a=0$ to get rid of the linear piece of the gauge variation, $\overline{\partial}\chi$. The contribution of the right-most diagram corresponds to the gauge variation of the $\tilde{J}_a \mathcal{B}^a_{\overline{z}}$ coupling
\begin{equation}
    \sum_{l,k \geq 0} \bigg( \frac{1}{2 \pi i} \bigg) \frac{1}{l! k!}\underset{\mathbb{C}}{\int} d^2 z f^a_{bc} \tilde{J}_a[l,k] \partial_{v^1}^l \partial_{v^2}^k (\nu^b \mathcal{A}^c_{\overline{z}} + \chi^b \mathcal{B}^c_{\overline{z}}).
\end{equation}
\begin{figure}[t]
    \centering
    \begin{subfigure}[b]{0.3\textwidth}
    \centering
    \includegraphics[scale=0.3]{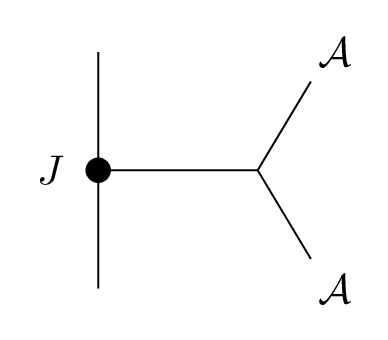}
    \end{subfigure}
    \begin{subfigure}[b]{0.3\textwidth}
\centering
         \includegraphics[scale=0.3]{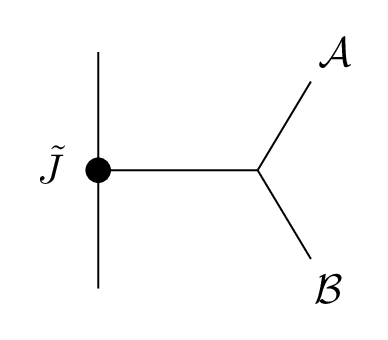}
    \end{subfigure}
     \caption{}
\end{figure}
Gauge invariance demands that these anomalous contributions be cancelled by the gauge variation of the diagrams in Figure 3. These contributions correspond to the linearized gauge variation of an integral involving two copies of the operators $J_a$, and an integral with one copy of $J_a$ and one of $\tilde{J}_a$, located at points $z$ and $w$ separated by a distance $|z-w| \geq \epsilon$, where $\epsilon$ is a small point-splitting regulator.\\ \\
The contribution from the left-most diagram in Figure 3 simplifies to 
\begin{equation}
    - \sum_{r,s,m,n \geq 0} \bigg( \frac{1}{2 \pi i} \bigg)^2 \frac{1}{r! s! m! n!}\underset{\mathbb{C}^2}{\int} d^2z d^2w (J_b [r,s](z) J_c [m,n](w)) \partial^r_{v^1} \partial^s_{v^2} \partial_{\overline{z}} \chi^b(z) \partial^m_{v^1} \partial^n_{v^2} \mathcal{A}^c_{\overline{z}}(w).
\end{equation}
Integrating by parts comes at the expense of introducing boundary terms where $|z-w| = \epsilon$. Taking the external legs to be test functions of the form $\chi = (v^1)^r (v^2)^s \mathbf{t}_c$ and $\mathcal{A}_{\overline{z}} = (v^1)^m (v^2)^n \mathbf{t}_b$, the requirement of cancellation becomes 
\begin{equation}
  \underset{\mathbb{C}}{\int} d^2 w f^a_{bc} J_a[r+m,s+n](w) = \underset{\mathbb{C}}{\int} d^2 w \underset{z \to w}{\text{Res}}(J_b[r,s](z)J_c[m,n](w)).
\end{equation}
Similarly, we find that cancellation of the second contribution leads to the following equality:
\begin{equation}
    \underset{\mathbb{C}}{\int} d^2 w f^a_{bc} \tilde{J}_a[r+m,s+n](w) = \underset{\mathbb{C}}{\int} d^2 w \underset{z \to w}{\text{Res}}(\tilde{J}_b[r,s](z)J_c[m,n](w)).
\end{equation}
This means that, at tree-level, gauge invariance of the coupling to the defect constrains the OPEs between the defect operators to be 
\begin{equation}
    \begin{aligned}
        J_b[r,s](z)J_c[m,n](w) \sim \frac{1}{z-w} f^a_{bc} J_a[r+m,s+n](w) \\
        \tilde{J}_b[r,s](z)J_c[m,n](w) \sim \frac{1}{z-w} f^a_{bc} \tilde{J}_a[r+m,s+n](w).
    \end{aligned}
\end{equation}
This is the level-$0$ Kac-Moody algebra for $\mathfrak{g}[v^1,v^2]$, discovered in the context of the celestial chiral algebra for gauge theory \cite{HOLOGRAPHIC_SYMMETRY_ALGEBRAS_FOR_GAUGE_THEORY_AND_GRAVITY}.
\section{Axion Field}
6d holomorphic BF theory suffers from an anomaly associated to the box diagram shown in Figure 4. This means that the 6d theory (and our chiral algebra) is not consistent at the quantum level. In 
\begin{figure}[t]
    \centering
    \begin{subfigure}[b]{0.3\textwidth}
    \centering
    \includegraphics[scale=0.3]{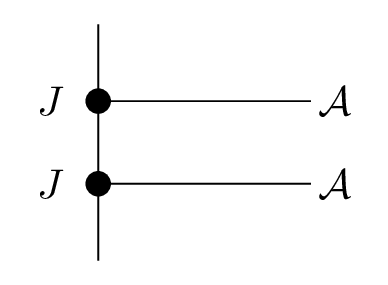}
    \end{subfigure}
    \begin{subfigure}[b]{0.3\textwidth}
\centering
         \includegraphics[scale=0.3]{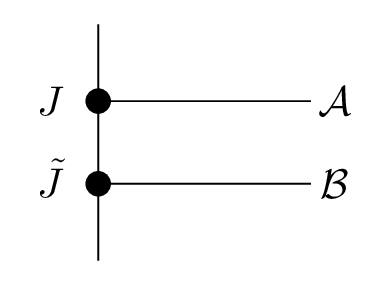}
    \end{subfigure}
     \caption{}
\end{figure}
\cite{QUANTIZING_HOLOMORPHIC_FIELD_THEORIES_ON_TWISTOR_SPACE}, it was demonstrated that this anomaly can be canceled by a Green-Schwarz mechanism under the condition that the gauge group is one of $\mathfrak{sl}_2$, $\mathfrak{sl}_3$, $\mathfrak{so}_8$ or one of the exceptional algebras.\\ \\
If the gauge group is one of the above, we extend our twistor action to include\footnote{The action becomes $S[\mathcal{A},\mathcal{B},\eta] = S[\mathcal{A},\mathcal{B}]+iS[\eta,\mathcal{A}]$}
\begin{equation}
    S[\eta,\mathcal{A}] = \frac{1}{2} \bigg( \frac{1}{2 \pi i}\bigg) \underset{\mathbb{PT}}{\int}\bigg( \partial^{-1} \eta \overline{\partial}\eta + \frac{\lambda_{\mathfrak{g}}}{(2 \pi i) \sqrt{12}} \eta \text{Tr}(\mathcal{A}\partial \mathcal{A}) \bigg)
\end{equation}
where we've enlarged our field content by introducing a new field $\eta \in \Omega^{2,1}(\mathbb{PT})$ constrained to satisfy $\partial \eta = 0$, and $\lambda_{\mathfrak{g}}$ is a constant that depends on the gauge group and is determined by the requirement that the gauge variation from Figure 5  (with the dashed line representing the $\eta$ propagator) cancels that of Figure 4.\footnote{The condition defining $\lambda_{\mathfrak{g}}$ is $\text{Tr}_{\text{adj}}(X^4) =  \lambda_{\mathfrak{g}}^2 \text{Tr}_{\text{fun}}(X^2)^2$.} This field is subject to the following gauge variation with generator $\gamma \in \Omega^{2,0}$
\begin{equation}
    \delta \eta = \overline{\partial} \gamma.
\end{equation}
Extending $\eta$ to a field in $\Omega^{2,*}(\mathbb{PT})[1]$, we find an additional term to the gauge variation of $\eta$ and $\mathcal{B}$
\begin{equation}
    \delta \eta = ...-\frac{\lambda_{\mathfrak{g}}}{(2 \pi i) \sqrt{12}} \text{Tr}(\partial \chi \partial \mathcal{A}) \quad \quad \delta \mathcal{B} = ...-\frac{i \lambda_{\mathfrak{g}}}{(2 \pi i) \sqrt{12}} (\gamma \partial \mathcal{A}+\eta \partial \chi).
\end{equation}
\\
In \cite{QUANTIZING_HOLOMORPHIC_FIELD_THEORIES_ON_TWISTOR_SPACE}, it was shown that this extension to the holomorphic BF theory is realized in 4d spacetime by a new axion-like field $\rho$\footnote{Here we are integrating over the $\mathbb{C}\mathbb{P}^{1}$ corresponding to  $x \in \mathbb{R}^4$.}
\begin{equation}
    \rho(x) = \frac{1}{2 \pi i} \underset{\mathbb{CP}^{1}_{x}}{\int} \partial^{-1} \eta,
\end{equation}
coupled to SDYM via
\begin{equation}
     \underset{\mathbb{R}^4}{\int} \bigg( \frac{1}{2}(\Delta\rho)^2 + \frac{\lambda_{\mathfrak{g}}}{(2 \pi i \sqrt{6})} \rho \text{Tr}(F(A) \wedge F(A)) \bigg).
\end{equation} 
\begin{figure}[t]
    \centering
    \begin{minipage}{0.3\textwidth}
    \centering
    \includegraphics[scale=0.35]{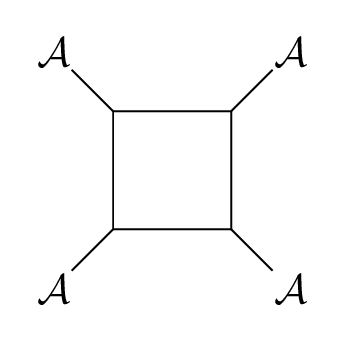}
     \caption{}
    \end{minipage} 
    \begin{minipage}{0.3\textwidth}
       \centering
    \includegraphics[scale=0.35]{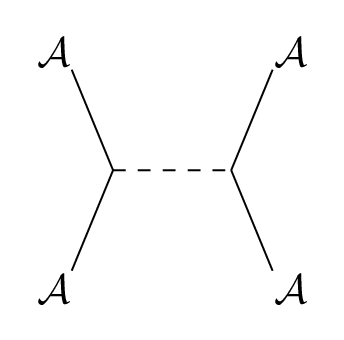}
     \caption{}
    \end{minipage}
\end{figure}
\section{Chiral Algebra Including The Axion}
The introduction of the axion field enlarges our chiral algebra by adding two extra towers  $E[r,s]$ and $F[r,s]$. These towers come from the defect coupling to the axion. Instead of working with $\eta$ directly, we choose to work with a (1,1)-form $\alpha$ satisfying $\partial \alpha = \eta$ to easily implement the constraint on $\eta$. This field is then subject to two gauge variations generated by $\omega \in \Omega^{0,1}(\mathbb{PT})$ and $\theta \in \Omega^{1,0}(\mathbb{PT})$:
\begin{equation}
    \delta \alpha = \partial \omega + \overline{\partial} \theta.
\end{equation}
The most general way that the defect theory couples to the axion is
\begin{equation}
     \sum_{m,n \geq 0} \bigg( \frac{1}{2 \pi i} \bigg) \frac{1}{m! n!} \underset{\mathbb{C}}{\int} d^2 z \bigg(e_i[m,n] \partial_{v^1}^m \partial_{v^2}^n \alpha_i + e_z[m,n] \partial_{v^1}^m \partial_{v^2}^n \alpha_z
    \bigg).
\end{equation}
Gauge invariance of the coupling under $\alpha \to \partial \omega$ leads to a constraint involving the $e_i$ and $e_z$ operators, which tells us that the operators are not independent. This is the reason why we only get two additional towers, as opposed to three. The towers $E[r,s]$ and $F[r,s]$ are then linear combinations of the $e_i$ and $e_z$ operators
\begin{equation}
    E[r,s]=\frac{-1}{r+s} e_z[r,s] \quad \quad F[r,s]=\frac{1}{r+s+2}(e_2[r+1,s]-e_1[r,s+1]).
\end{equation}
Using Koszul duality considerations similar to section 4, we can derive the tree-level OPEs of the enlarged chiral algebra. For more details, see section 7.2 of \cite{CELESTIAL_HOLOGRAPHY_MEETS_TWISTED_HOLOGRAPHY_4D_AMPLITUDES_FROM_CHIRAL_CORRELATORS} and section 10 of \cite{TWISTED_HOLOGRAPHY_AND_CELESTIAL_HOLOGRAPHY_FROM_BOUNDARY_CHIRAL_ALGEBRA}.
\section{One-Loop Corrections}
 In \cite{ON_THE_ASSOCIATIVITY_OF_ONE_LOOP_CORRECTIONS_TO_THE_CELESTIAL_OPE}, the quantum corrections to the $J[0,1]J[1,0]$, $\tilde{J}[0,1] J[1,0]$, and $\tilde{J}[1,0] J[0,1]$ OPEs were computed using known 4d collinear splitting amplitudes and constraints from associativity. Here, we demonstrate how the same result can be obtained through Koszul duality using the methods developed in \cite{KOSZUL_DUALITY_IN_QFT,ON_THE_ASSOCIATIVITY_OF_1_LOOP_CORRECTIONS_TO_THE_CELESTIAL_OPERATOR_PRODUCT_IN_GRAVITY}. In particular, a similar computation in the setting of self-dual gravity coupled to a $4^{\text{th}}$-order gravitational axion was performed in \cite{ON_THE_ASSOCIATIVITY_OF_1_LOOP_CORRECTIONS_TO_THE_CELESTIAL_OPERATOR_PRODUCT_IN_GRAVITY}, and we closely follow the presentation therein. 
 \begin{figure}[t]
    \centering
    \begin{subfigure}[b]{0.3\textwidth}
    \centering
    \includegraphics[scale=0.35]{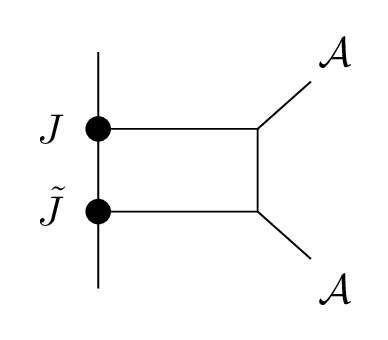}
    \end{subfigure}
    \begin{subfigure}[b]{0.3\textwidth}
\centering
         \includegraphics[scale=0.35]{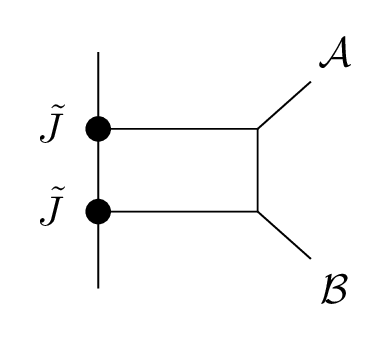}
    \end{subfigure}
     \caption{}
     \label{figure_1}
\end{figure}
\\ \\
Since axion exchanges are already counted as a one-loop effect by the Green-Schwarz mechanism, one-loop corrections to these OPEs cannot involve the axion operators.\footnote{In our counting convention, any $E$ or $F$ operator on the right-hand side of an OPE comes with an explicit factor of $\hbar$. Since any one-loop effect also comes with an explicit factor of $\hbar$, it follows that the 1-loop corrections cannot involve $E$ or $F$.} Reintroducing $\hbar$, SDYM is invariant under simultaneous rescalings $\hbar \to \lambda \hbar$, $\mathcal{B} \to \lambda \mathcal{B}$. In terms of the chiral algebra, this means that under this rescaling, $\tilde{J}$ transforms non-trivially, $\tilde{J} \to \lambda^{-1} \tilde{J}$. 
Since 1-loop corrections come with an explicit factor of $\hbar$, invariance under rescaling of $\hbar$ requires that the number of $\tilde{J}$'s increases by one so that the powers of $\lambda$ match on both sides of the OPE. \\ \\
The one-loop Feynman diagrams that yield anomalies in the bulk-defect coupling which contribute corrections are then illustrated in Figure 6. A priori, we should also consider the BRST variation of the diagram in Figure 7. In \cite{RENORMALIZATION_FOR_HOLOMORPHIC_FIELD_THEORIES} it was proven that this contribution necessarily vanishes in a 6d holomorphic theory.
We reproduce the corrections to the $\tilde{J}[0,1] J[1,0]$, and $\tilde{J}[1,0] J[0,1]$ OPEs in Appendix B. 
\\ \\
We denote the location of the defect operators as $Z = (z_1,0^{\Dot{\alpha}})$ and $W = (z_2,0^{\Dot{\alpha}})$, where we again require that their distance satisfy $|z_1-z_2| \geq \epsilon$. We denote the location of the vertices as $X = (x^0,x^{\Dot{\alpha}})$ and $Y = (y^0,y^{\Dot{\alpha}})$. We also define $z_0 = \frac{z_1+z_2}{2}$, $z_{12}=z_1-z_2$, and $D^{i}_{rs} = \frac{1}{r! s!} \partial^{r}_{v^{1}_{i}} \partial^{s}_{v^{2}_{i}}$. \\ \\
The linearized BRST variation of the left-most diagram in Figure 6 leads to two terms. The term corresponding to the gauge variation acting on the up-most external leg is given by the general form
\begin{equation}
     \bigg( \frac{1}{2 \pi i} \bigg)^2 \underset{\mathbb{C}^2}{\int} dz_1 dz_2  (J_c[r,s](z_1) \tilde{J}_d[m,n](z_2)) K^{ef}f^d_{ae} f^c_{bf} \underset{rsmn}{\mathcal{M}}(z_1,z_2; \overline{\partial} \chi^b, \mathcal{A}^a) 
\end{equation}
\begin{equation}
    \underset{rsmn}{\mathcal{M}}(z_1,z_2; \overline{\partial}\chi^b, \mathcal{A}^a) = 
    \bigg(\frac{1}{2}\bigg)^2 \bigg( \frac{1}{2 \pi i} \bigg)^2 \underset{(\mathbb{C}^3)^2}{\int} D^1_{rs}\underset{z \leftrightarrow x}{P} \overline{\partial} \chi^{b} d^3X \underset{x \leftrightarrow y}{P} d^3Y \mathcal{A}^{a} D^4_{mn}\underset{y \leftrightarrow w}{P} 
\end{equation}
where the structure constants $f^a_{bc}$ come from the trivalent vertices labeled by the cubic interaction of the action, the Killing form $K^{fe}$ come from the bivalent vertex labeled by the quadratic interaction, and $\underset{z \leftrightarrow w}{P}$ is the $\mathcal{A}-\mathcal{B}$ propagator without the Lie algebra information. This propagator is defined via the equations 
 \begin{equation}
     \underset{z \leftrightarrow w}{P} = j^{*}(p), \quad \quad \frac{1}{2 \pi i} \overline{\partial}p = -\delta^{(3)}, \quad \quad \underset{\mathbb{C}^3}{\int} d^3Z\delta^{(3)} = 1  
 \end{equation}
 where $j:(\mathbb{C}^3 \times \mathbb{C}^3 \to \mathbb{C}^3)$ is the difference map $(Z,W) \mapsto (Z-W)$, $p \in \Omega^{(0,2)}(\mathbb{C}^3)$, and $\delta^{(3)}$ is the $(0,3)$-form $\delta$-function with support at $Z=0$. A nice discussion of propagators and Feynman rules in holomorphic gauge theories can be found in \cite{A_ONE_LOOP_EXACT_QUANTIZATION_OF_CHERN_SIMONS_THEORY}.  Explicitly, the propagator is given by
\begin{equation}
    \underset{z \leftrightarrow w}{P} = \bigg(\frac{1}{2 \pi}\bigg)^2 \epsilon_{\overline{a} \overline{b} \overline{c}} \frac{(\overline{Z}-\overline{W})^{\overline{a}}d(\overline{Z}-\overline{W})^{\overline{b}}d(\overline{Z}-\overline{W})^{\overline{c}}}{\lVert Z-W \rVert^6}.
\end{equation} \\
The second term, which corresponds to the gauge variation acting on the down-most external leg, can be obtained from the first by exchanging $J \leftrightarrow \tilde{J}$ before taking any OPEs. \\ \\
Integrating by parts leads to terms where $\overline{\partial}$ acts on a propagator, and a boundary term where $\lvert z_{12} \rvert = \epsilon$. Since $\underset{z \leftrightarrow w}{\overline{\partial}{P}} = -(2 \pi i) \underset{z \leftrightarrow w}{\delta^{(3)}}$, where $\underset{z \leftrightarrow w}{\delta^{(3)}}$ is the $(0,3)$-form $\delta$-function with support at $Z-W=0$, these terms correspond to contractions of the internal edges. Diagrammatically this takes the form of Figure 8. 
\begin{figure}[t]
    \centering
    \includegraphics[scale=0.35]{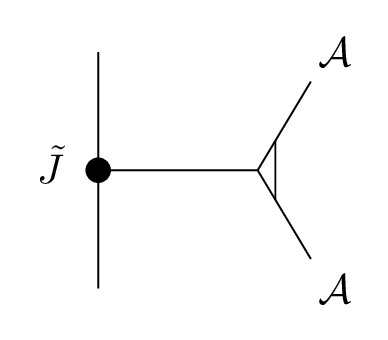}
    \caption{}
\end{figure}
 \\ \\
Consider the contraction coming from the term with $\underset{z_{i} \leftrightarrow z_{j}}{\overline{\partial} P}$. The resulting expression will contain holomorphic derivatives acting on the product $\underset{z_1 \leftrightarrow z_k }{P} \chi \mathcal{A} \underset{z_k \leftrightarrow z_4}{P}$. This product will be proportional to:
\begin{equation}
    (\epsilon_{\overline{a}\overline{b}\overline{c}}  \overline{Z}^{\overline{a}}_{{1k}} d\overline{Z}^{\overline{b}}_{{1k}} d\overline{Z}^{\overline{c}}_{{1k}})(  \epsilon_{\overline{d}\overline{e}\overline{f}}  \overline{Z}^{\overline{d}}_{{k2}} d\overline{Z}^{\overline{e}}_{{k2}} d\overline{Z}^{\overline{f}}_{{k2}}).
\end{equation} \\
Denoting the location of the vertex as $Z_{k} = (z_k,v^{\dot{\alpha}}_{k})$, the differences $Z_{1k}$ and $Z_{k2}$ become $(z_{1k},-v^{\dot{\alpha}}_{k})$ and $(z_{k2},v^{\dot{\alpha}}_{k})$. The only terms that survive in eq. 7.5 are those which do not have more than two $d\overline{v}^{\dot{\alpha}}_{k}$,
\begin{equation}
    (\epsilon_{\overline{a}\overline{b}\overline{c}}  \overline{Z}^{\overline{a}}_{{1k}} d\overline{Z}^{\overline{b}}_{{1k}} d\overline{Z}^{\overline{c}}_{{1k}})(  \epsilon_{\overline{d}\overline{e}\overline{f}}  \overline{Z}^{\overline{d}}_{{k2}} d\overline{Z}^{\overline{e}}_{{k2}} d\overline{Z}^{\overline{f}}_{{k2}}) = -4\overline{v}^{\dot{\alpha}} \overline{v}^{\dot{\beta}} \epsilon_{\dot{\alpha} \dot{\gamma}} \epsilon_{\dot{\beta} \dot{\nu}} \epsilon^{\dot{\nu}\dot{\gamma}} d^{2}\overline{v}_{k}d\overline{z}_{1k} d\overline{z}_{k2}.
\end{equation}
 \begin{figure}[t]
     \centering
     \includegraphics[scale=0.35]{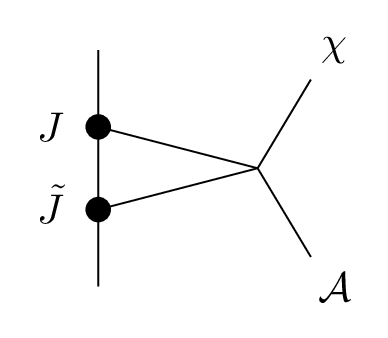}
     \caption{}
 \end{figure} 
 Summing over the $\epsilon$ indices, we find that the contributions coming from contractions are proportional to $\epsilon_{\dot{\alpha} \dot{\beta}}\overline{v}^{\dot{\alpha}} \overline{v}^{\dot{\beta}} = [\overline{v},\overline{v}] = 0$. This means that the only anomalous contributions come from the boundary term, which takes the form:
\begin{equation}
    -\bigg( \frac{1}{2 \pi i} \bigg)^2 \underset{\mathbb{C}}{\int} dz_0 \underset{|z_{12}| = \epsilon}{\oint} dz_{12} (J_c[r,s](z_1) \tilde{J}_d[m,n](z_2)) K^{ef}f^d_{ae} f^c_{bf} \underset{rsmn}{\mathcal{M}}(z_1,z_2; \chi^b, \mathcal{A}^a ) 
\end{equation}
where we restrict $d\overline{z}_1 = d\overline{z}_2 = d\overline{z}_0$. As in section 4, gauge invariance requires that this be cancelled by the gauge variation of the left-most diagram in Figure 3
\begin{equation}
     -\bigg( \frac{1}{2 \pi i} \bigg)^2 \underset{\mathbb{C}^2}{\int} d^2z_{1} d^2z_{2} (J_b [0,1](z_1) J_a [1,0](z_2)) D^1_{01} \partial_{\overline{z}}\chi^b(z_1) D^2_{10} \mathcal{A}^a_{\overline{z}}(z_2).
\end{equation}
The matching of scaling dimensions then tells us then that the only terms that contribute in eq. 7.7 are those which satisfy $r+s+m+n=0$. This follows from the fact that $J[0,1]J[1,0]$ has scaling dimension equal to $-2$. We therefore need to compute the following quantity:
\begin{equation}
     \underset{0000}{\mathcal{M}}(z_1,z_2; \chi^b, \mathcal{A}^a) = \bigg(\frac{1}{2}\bigg)^2 \bigg( \frac{1}{2 \pi i} \bigg)^2 \underset{(\mathbb{C}^3)^2}{\int} \underset{z \leftrightarrow x}{P} \chi^b d^3X \underset{x \leftrightarrow y}{P} d^3Y \mathcal{A}^a \underset{y \leftrightarrow w}{P} \bigg|_{d\overline{z}_1 = d\overline{z}_2 = d\overline{z}_0}.
\end{equation}
 We take the external legs to be test functions of the form $\chi= z(v^2) \mathbf{t}_b$ and $\mathcal{A} = (v^1) d\overline{z} \mathbf{t}_a$:
\begin{equation}
      \underset{0000}{\mathcal{M}}(z_1,z_2; x^0 x^2, y^1 d\overline{y}^0) = -\bigg(\frac{1}{2 \pi i}\bigg)^2  \bigg(\frac{1}{4 \pi^2}\bigg)^3 \bigg( \frac{1}{2} \bigg)^2 8 \overline{z}_{12} d\overline{z}_0 \underset{(\mathbb{C}^3)^2}{\mathcal{I}} \end{equation}
      \begin{equation}
      \underset{(\mathbb{C}^3)^2}{\mathcal{I}} = \underset{(\mathbb{C}^3)^2}{\int} d^6X d^6Y \frac{x^0 [\overline{x},\overline{y}] x^2 y^1}{\lVert Z-X \rVert^6 \lVert X-Y \rVert^6 \lVert Y-W \rVert^6}.
\end{equation}
where we've used the fact that the antiholomorphic form structure
\begin{equation}
      (\epsilon_{\overline{a}\overline{b}\overline{c}}  \overline{Z}^{\overline{a}}_{{1x}} d\overline{Z}^{\overline{b}}_{{1x}} d\overline{Z}^{\overline{c}}_{{1x}})(  \epsilon_{\overline{d}\overline{e}\overline{f}}  \overline{Z}^{\overline{d}}_{{xy}} d\overline{Z}^{\overline{e}}_{{xy}} d\overline{Z}^{\overline{f}}_{{xy}})(  \epsilon_{\overline{g}\overline{h}\overline{i}}  \overline{Z}^{\overline{g}}_{{y2}} d\overline{Z}^{\overline{h}}_{{y2}} d\overline{Z}^{\overline{i}}_{{y2}}). 
\end{equation}
with $d\overline{z}_1 = d\overline{z}_2 = d\overline{z}_0$, simplifies to:
\begin{equation}
    8 [\overline{v}_{x},\overline{v}_{y}] \overline{z}_{12} d\overline{z}_0 d\overline{z}_x d^2\overline{v}_{x} d^2 \overline{v}_{y}.
\end{equation}
We compute this integral explicitly in the appendix. The result is 
\begin{equation}
\underset{(\mathbb{C}^3)^2}{\mathcal{I}} = -\frac{(-2 \pi i)^6}{(2!)^3 6} \frac{3z_0 + \frac{z_{12}}{2}}{|z_{12}|^2} \quad \quad 
       \underset{0000}{\mathcal{M}}(z_1,z_2; x^0 x^2, y^1 d\overline{y}^0) = \frac{1}{96 \pi^2} \bigg(\frac{3z_0}{z_{12}} + \frac{1}{2}\bigg) d\overline{z}_0.
\end{equation}
After inserting this into eq. 7.7 and performing the contour integral, the anomalous contribution coming from both terms then becomes
\begin{equation}
    \bigg(\frac{1}{2 \pi i}\bigg) \frac{1}{96 \pi^2} \underset{\mathbb{C}}{\int} d^2z_0 \bigg(3 z_{0} C^{cd}_{ab} :J_c[0,0] \tilde{J}_d[0,0]:(z_0)+ K^{ef}f^{d}_{ae}f^{c}_{bf}f^{l}_{cd} \tilde{J}_l[0,0](z_0)\bigg)
\end{equation}
\begin{equation*}
     C^{cd}_{ab}= K^{ef}(f^d_{ae} f^c_{bf}+f^c_{ae} f^d_{bf}).
\end{equation*}
We can simplify the second term by repeated use of the Jacobi identity after writing it as
\begin{equation}
    K^{ef}f^d_{ae} f^c_{bf} f^l_{dc} = \frac{1}{2}K^{ef}(f^d_{ae}f^c_{bf}-f^d_{be}f^c_{af})f^l_{dc}
\end{equation}
and making use of the fact that the Casimir in the adjoint representation is $2h^{\vee}$, which gives us the identity:
\begin{equation}
    K^{ab}f^e_{ad}f^d_{bc}= 2h^{\vee} \delta^e_c.
\end{equation}
We obtain
\begin{equation}
     \bigg(\frac{1}{2 \pi i}\bigg) \underset{\mathbb{C}}{\int} \frac{d^2z_0}{96 \pi^2} \bigg(3z_0K^{ef}(f^c_{ae} f^d_{bf}+f^d_{ae} f^c_{bf}) :J_c[0,0] \tilde{J}_d[0,0]:(z_0)+h^{\vee} f^c_{ab} \tilde{J}_c[0,0](z_0)\bigg).
\end{equation} 
With this choice of test functions, eq. 7.8 simplifies to:
\begin{equation}
    -\bigg(\frac{1}{2 \pi i} \bigg)^2 \underset{\mathbb{C}}{\int} d^2z_0 \underset{|z_{12}| = \epsilon}{\oint} dz_{12} (J_b[0,1](z_1)J_a[1,0](z_2))(\frac{z_{12}}{2}+z_0).
\end{equation}
We need to perform the contour integral.
\begin{equation}
    -\bigg( \frac{1}{2 \pi i} \bigg) \underset{\mathbb{C}}{\int} d^2z_0 \underset{z_1 \to z_2}{\text{Res}}\bigg( (\frac{z_{12}}{2}+z_0) (J_b[0,1](z_1)J_a[1,0](z_2)) \bigg).
\end{equation}
Comparing eq. 7.20 to eq. 7.18, we see that gauge invariance holds if and only if the OPE correction has the form
\begin{equation}
\begin{aligned}
    J_b[0,1](z)J_a[1,0](w) \sim \frac{\alpha}{z-w} K^{ef}(f^c_{ae} f^d_{bf}+f^d_{ae} f^c_{bf}) :J_c[0,0] \tilde{J}_d[0,0]:(w) \\
    +\frac{\beta}{(z-w)^2} f^c_{ab} \tilde{J}_c[0,0](w)+\frac{\beta}{2(z-w)}f^c_{ab}\partial\tilde{J}_c[0,0](w),
    \end{aligned}
\end{equation}
with the numerical constants \footnote{This matches what was found in \cite{ON_THE_ASSOCIATIVITY_OF_ONE_LOOP_CORRECTIONS_TO_THE_CELESTIAL_OPE} after adjusting the normalization convention. See footnote \ref{footnote_1}.}
\begin{equation}
    \alpha = \frac{1}{32 \pi^2} = -\frac{3}{2(2\pi i)^2 12} \quad \quad \beta = \frac{h^{\vee}}{48 \pi^2} = - \frac{h^{\vee}}{(2 \pi i)^2 12}.
\end{equation}
Note that the added term $\partial \tilde{J}$ is a consequence of symmetry, in particular, the requirement that
\begin{equation}
    J_b[1,0](z)J_a[0,1](w) \sim - J_a[0,1](z)J_b[1,0](w).
\end{equation}
Similarly, we find in Appendix B that the $\tilde{J}[0,1]J[1,0]$ and $\tilde{J}[1,0] J[0,1]$ OPEs receive the following corrections:
\begin{equation}
\begin{aligned}
    \tilde{J}_b[0,1](z) J_a[1,0](w) \sim \frac{\alpha}{z-w} K^{fe}f^c_{ae}f^d_{bf}:\tilde{J}_c[0,0]\tilde{J}_d[0,0]:(w) \\
     \tilde{J}_b[1,0](z) J_a[0,1](w) \sim \frac{-\alpha}{z-w} K^{fe}f^c_{ae}f^d_{bf}:\tilde{J}_c[0,0]\tilde{J}_d[0,0]:(w).
     \end{aligned}
\end{equation}
Notice that even though the axion did not appear explicitly in the computation of the one-loop diagram (in contrast to the computations in \cite{ON_THE_ASSOCIATIVITY_OF_ONE_LOOP_CORRECTIONS_TO_THE_CELESTIAL_OPE} which make direct use of the extended tree-level OPEs, including the $E$ and $F$ generators) Koszul duality is guaranteed to output a well-defined associative chiral algebra, with the precise numerical coefficients characteristic of the axion-coupled twistor theory.\footnote{We focused on the corrections to the OPEs involving $J[1]$ since $J[1]$ generate the chiral algebra. Similarly, we can also compute the one-loop deformations of OPEs between $J[k]J[1]$ and $\tilde{J}[k] J[1]$ by adjusting the external legs test-functions and using arguments of dimension-matching, for more details see Appendix C. In principle, we expect that these can also be determined from the $J[1]$ OPEs and the requirement of associativity. Here, $J[k]$ denotes $J[k_1,k_2]$ with $k_1+k_2 = k$.}

\acknowledgments
I would like to thank Natalie Paquette, my research advisor, for assigning me this project. I am deeply grateful for her continued support, instruction and exceptional guidance. I would also like to thank Niklas Garner for helpful discussions and invaluable comments on a draft version of this paper. V.F. acknowledges support from the University of Washington and the U.S. Department of Energy Early Career Award, DOE award DE-SC0022347.

\appendix
\section{Explicit Integral Calculation}
This calculation is a slight adaptation of the integral techniques employed in Appendix C of \cite{ON_THE_ASSOCIATIVITY_OF_1_LOOP_CORRECTIONS_TO_THE_CELESTIAL_OPERATOR_PRODUCT_IN_GRAVITY}, whose notation we largely follow. We spell out the details below. \\ \\
We first integrate over $d^6Y$. Using Feynman parametrization, 
\begin{equation}
   \overset{n}{\underset{i=1}{\Pi}} \frac{1}{c^{\alpha_i}_i} = \frac{(\overset{n}{\underset{i=1}{\sum}}c^{\alpha_i}_i-1)!}{ \overset{n}{\underset{i=1}{\Pi}}(c^{\alpha_i}_i-1)!} \underset{[0,1]}{\int} \overset{n}{\underset{i=1}{\Pi}} dt_i t_i^{\alpha_i-1} \delta(1-\overset{n}{\underset{i=1}{\sum}}t_i)(\overset{n}{\underset{i=1}{\sum}}t_i c_i)^{-\overset{n}{\underset{i=1}{\sum}}\alpha_i},
\end{equation}
we can rewrite the integral as
\begin{equation}
     \underset{\mathbb{C}^3}{\int} d^6X\frac{5!}{(2!)^2} \frac{1}{\lVert Z-X \rVert^6} \underset{[0,1]}{\int} dt t^2(1-t)^2 \underset{\mathbb{C}^3}{\int} d^6Y \frac{x^0 [\overline{x},\overline{y}] x^2y^1}{(t \lVert X-Y \rVert^2 + (1-t) \lVert Y-W \rVert^2)^6}.
\end{equation}
We define $\tilde{Y} = Y-tX-(1-t)W$ and retain only the terms that are invariant under phase rotations of $\tilde{y}^{\dot{\alpha}}$. These are the only terms that have nonvanishing contributions:
\begin{equation}
    -\underset{\mathbb{C}^3}{\int} d^6X\frac{5!}{(2!)^2} \frac{1}{\lVert Z-X \rVert^6} \underset{[0,1]}{\int} dt t^2(1-t)^2 \underset{\mathbb{C}^3}{\int} d^6\tilde{Y} \frac{x^0  |x^2|^2 |\tilde{y}^1|^2}{(\lVert \tilde{Y} \rVert^2 + t(1-t) \lVert X-W \rVert^2)^6}.
\end{equation}
Now we define $r_i = |\tilde{y}^i|^2$ and use $d^2\tilde{y}^{i} = -idr_{i}d\theta_{i}$ to simplify the above integral to
\begin{equation}
     -\underset{\mathbb{C}^3}{\int} d^6X\frac{5!(-2\pi i)^3 x^0 |x^2|^2}{(2!)^2 \lVert Z-X \rVert^6}  \underset{[0,1]}{\int} dt t^2(1-t)^2 \underset{[0,\infty)^3}{\int} dr_0 dr_1 dr_2 \frac{r_2}{(\overset{2}{\underset{i=0}{\sum}}r_i+t(1-t)\lVert X-W \rVert^2)^{6}}.
\end{equation}
Using eq. A.1 with $c_i =1$, we can integrate over $dr_0 dr_1 dr_2$. We now perform the same steps as before to integrate over $d^6X$. We define $\tilde{X} = X -tZ -(1-t)W$, and retain only the terms that are invariant under phase rotations of $x^{0}$:
\begin{equation}
   - \frac{4!(-2 \pi i)^3}{(2!)^3} \underset{[0,1]}{\int} dt t^2(1-t) (tz_1 +(1-t)z_2)\underset{\mathbb{C}^3}{\int} d^6\tilde{X} \frac{|\tilde{x}^2|^2}{(\lVert \tilde{X} \rVert^2+t(1-t)|z_1-z_2|^2)^5}.
\end{equation}
Defining $r_i = |\tilde{x}^i|^2$ and integrating over $dr_0 dr_1 dr_2 d\theta_0 d\theta_1 d\theta_2$, we are left with
\begin{equation}
    -\frac{(-2 \pi i)^6}{(2!)^3 |z_1-z_2|^2} \underset{[0,1]}{\int} dt (t^2z_1+t(1-t)z_2).
\end{equation}
Performing the integral and rewriting the resulting expression in terms of $z_{12}$ and $z_0$, we finally obtain 
\begin{equation}
    \underset{(\mathbb{C}^3)^2}{\mathcal{I}} =- \frac{(-2 \pi i)^6}{(2!)^3 6} \frac{3z_0 + \frac{z_{12}}{2}}{|z_{12}|^2}.
\end{equation}
\section{Remaining Anomalies}
The remaining anomalous contributions lead to quantum corrections of the $\tilde{J}[0,1] J[1,0]$ and $\tilde{J}[1,0] J[0,1]$ OPEs. In \cite{ON_THE_ASSOCIATIVITY_OF_1_LOOP_CORRECTIONS_TO_THE_CELESTIAL_OPERATOR_PRODUCT_IN_GRAVITY}, it was determined that these corrections were entirely determined by the numerical constants we've calculated. In this section of the Appendix, we will show this to be the case using the methods of section 7. \\ \\ The scaling dimension of $\tilde{J}[0,1] J[1,0]$ and $\tilde{J}[1,0] J[0,1]$ is $-4$, meaning that the one-loop contribution only comes from the term with $(m,n,r,s) = (0,0,0,0)$. To obtain the $\tilde{J}[0,1] J[1,0]$ OPE correction, we take the external legs to be test functions of the form $\nu = z(v^2) \mathbf{t}_b$ and $\mathcal{A} = (v^1) d\overline{z} \mathbf{t}_a$.\footnote{It is sufficient to only consider the linearized-BRST gauge variation acting on $\mathcal{B}$.} The one-loop contribution is then
\begin{equation}
    -\bigg( \frac{1}{2 \pi i}\bigg)^2 \underset{\mathbb{C}}{\int} d z_0 \underset{|z_{12}| =\epsilon}{\oint}dz_{12} \tilde{J}_c[0,0](z_1) \tilde{J}_d[0,0](z_2) K^{fe}f^d_{ae}f^c_{bf}  \underset{0000}{\mathcal{M}}(z_1,z_2; x^0 x^2, y^1 d\overline{y}^0).
\end{equation}
Using the techniques from Appendix A, this becomes
\begin{equation}
    \bigg(\frac{1}{2 \pi i}\bigg) \underset{\mathbb{C}}{\int} d z_0 \alpha z_0K^{fe}f^d_{ae}f^c_{bf}:\tilde{J}_c[0,0]\tilde{J}_d[0,0]:(z_0).
\end{equation}
The gauge variation of the right-most diagram of Figure 3 contributes
\begin{equation}
    -\bigg( \frac{1}{2 \pi i}\bigg)^2 \underset{\mathbb{C}}{\int} dz_0 \underset{|z_{12}| =\epsilon}{\oint}dz_{12}(\tilde{J}_b[0,1](z_1)J_a[1,0](z_2))(\frac{z_{12}}{2}+z_0)).
\end{equation}
It readily follows that gauge invariance holds only if the corrections to the OPE are given by
\begin{equation}
    \tilde{J}_b[0,1](z) J_a[1,0](w) \sim \frac{\alpha}{z-w} K^{fe}f^c_{ae}f^d_{bf}:\tilde{J}_c[0,0]\tilde{J}_d[0,0]:(w).
\end{equation} \\
For the $\tilde{J}[1,0] J[0,1]$ OPE correction, we take the external legs to be test functions of the form $\nu = z(v^1) \mathbf{t}_b$ and $\mathcal{A} = (v^2) d\overline{z} \mathbf{t}_a$. Using the fact that $\underset{0000}{\mathcal{M}}(z_1,z_2; x^0 x^1, y^2 d\overline{y}^0) = -\underset{0000}{\mathcal{M}}(z_1,z_2; x^0 x^2, y^1 d\overline{y}^0)$, we find that the corrections to the OPE must be
\begin{equation}
    \tilde{J}_b[1,0](z) J_a[0,1](w) \sim \frac{-\alpha}{z-w} K^{fe}f^c_{ae}f^d_{bf}:\tilde{J}_c[0,0]\tilde{J}_d[0,0]:(w).
\end{equation}

\section{General Form of the 1-Loop Corrections}
With an appropriate choice of test functions for the external legs, the coefficients of the 1-loop corrections are encoded in the $\mathcal{M}$ "scattering" elements.  Let the test functions be of the form:
\begin{equation}
    \chi = z(x^1)^r(x^2)^s\mathbf{t}_b, \quad \quad \mathcal{A} = (y^1)^m(y^2)^nd\overline{y}^0 \mathbf{t}_a.
\end{equation}
With this choice of test functions, $\mathcal{M}$ is given by:
\begin{equation}
      \underset{k_1 k_2 l_1 l_2}{\mathcal{M}} = -\bigg( \frac{1}{2} \bigg)^2 \bigg(\frac{1}{2\pi i}\bigg)^2 \bigg(\frac{1}{4 \pi^2} \bigg)^3 8\overline{z}_{12}d\overline{z}_0 \frac{(2+k_1+k_2)!(2+l_1+l_2)!}{2^2k_1!k_2!l_1!l_2!}\underset{(\mathbb{C}^3)^2}{\mathcal{I}} \\
\end{equation} \\
\begin{equation}
     \underset{(\mathbb{C}^3)^2}{\mathcal{I}} = \underset{(\mathbb{C}^3)^2}{\int} d^6X d^6Y \frac{x^0 [\overline{x},\overline{y}] (x^1)^{r}(\overline{x}^1)^{k_1}(x^2)^{s}(\overline{x}^2)^{k_2}(y^1)^{m}(\overline{y}^1)^{l_1}(y^2)^n (\overline{y}^2)^{l_2}}{(\lVert Z-X \rVert^2)^{3+k_1+k_2} (\lVert X-Y \rVert^2)^3 (\lVert Y-W \rVert^2)^{3+l_1+l_2}}.
\end{equation} \\
Using Feynman parametrization and defining $\tilde{Y}=Y-tX-(1-t)W$, the integral over $d^6Y$ becomes:
\begin{equation*}
    \frac{(5+l_1+l_2)!}{2(2+l_1+l_2)!}\underset{[0,1]}{\int}dt(1-t)^{2+l_1+l_2}t^2 \underset{\mathbb{C}^3}{\int}d^6\tilde{Y}\frac{[\overline{x},\overline{\tilde{y}}](\tilde{y}^1+tx^1)^{m}(\overline{\tilde{y}^1}+t\overline{x}^1)^{l_1}(\tilde{y}^2+tx^2)^n (\overline{\tilde{y}^2}+t\overline{x}^2)^{l_2}}{(\lVert \tilde{Y}\rVert^2+t(1-t) \lVert X-W \rVert^2)^{6+l_1+l_2}}.
\end{equation*}
We need to only retain the terms which are invariant under phase rotations of $\tilde{y}^{\dot{\alpha}}$. Using the binomial theorem, the numerator of the integrand is:
\begin{equation}
\underset{abcd}{\sum}(\overline{x}^1 \overline{\tilde{y}^2} - \overline{x}^2 \overline{\tilde{y}^1})(\tilde{y}^1)^{a}(tx^1)^{m-a}(\overline{\tilde{y}^1})^{b}(t\overline{x}_1)^{l_1-b}(\tilde{y}^2)^{c}(tx^2)^{n-c}(\overline{\tilde{y}^2})^{d}(t\overline{x}^2)^{l_2-d}
\end{equation} 
where we've defined
\begin{equation}
    \underset{abcd}{\sum} =\sum_{a=0}^{m}\sum_{b=0}^{l_1}\sum_{c=0}^{n}\sum_{d=0}^{l_2}{m \choose a}{l_1 \choose b}{n \choose c}{l_2 \choose d}.
\end{equation}
The only terms that contribute from the first piece of eq. C.4 are those satisfying $a=b$ and $c=d+1$, and the only terms that contribute from the second piece are those satisfying $a=b+1$ and $c=d$. Defining
\begin{equation*}
    \underset{ac}{\sum} = \bigg(\sum_{a=0}^{\text{min}(m,l_1)} \sum_{c=1}^{\text{min}(n,l_2+1)} {l_1 \choose a}{l_2 \choose c-1} - \sum_{a=1}^{\text{min}(m,l_1+1)} \sum_{c=0}^{\text{min}(n,l_2)} {l_1 \choose a-1}{l_2 \choose c}\bigg) {m \choose a} {n \choose c}
\end{equation*}
and $r_i = \lvert \tilde{y}^i \rvert^2$, we can perform the integrals and obtain 
\begin{equation}
    \underset{ac}{\sum} \frac{a! c! (m+n-a-c)!(a+c-1)!(2+l_1+l_2-a-c)!}{2(2+l_1+l_2)!(m+n)!} (-2 \pi i)^3  \underset{\mathbb{C}^3}{\mathcal{I}}
\end{equation}
where $\underset{\mathbb{C}^3}{\mathcal{I}}$ corresponds to the integral over $d^6X$ with the added $x^{\dot{\alpha}}$ terms coming from the integration over $d^6Y$:
\begin{equation}
\underset{\mathbb{C}^3}{\mathcal{I}} = \underset{\mathbb{C}^3}{\int} d^6X
\frac{x^0 (x^1)^{r+m-a}(\overline{x}^1)^{k_1+l_1+1-a}(x^2)^{s+n-c}(\overline{x}^2)^{k_2+l_2+1-c}}{(\lVert Z-X \rVert^2)^{3+k_1+k_2}(\lVert X-W \rVert^2)^{3+l_1+l_2-a-c}}.
\end{equation}
To integrate over $d^6X$, we perform the same steps as above. Namely, we use Feynman parametrization, then define $\tilde{X}=X-tZ-(1-t)W$, and only retain the terms which are invariant under phase rotations of $x^0$ and $x^{\dot{\alpha}}$. We find that \\
\begin{equation}
    \underset{\mathbb{C}^3}{\mathcal{I}} =\frac{(-2 \pi i)^3}{\lvert z_1-z_2 \rvert^2} \frac{(r+m-a)!(s+n-c)! C(z_1,z_2)}{(2+k_1+k_2)!(2+l_1+l_2-a-c)!(2+m+r+n+s-a-c)!}
\end{equation} \\
\begin{equation}
   C(z_1,z_2) = \bigg( (2+k_1+k_2)!(1+l_1+l_2-a-c)!z_1+ (1+k_1+k_2)!(2+l_1+l_2-a-c)!z_2\bigg)
\end{equation} \\
and the requirement that $m+r=k_1+l_1+1$ and $n+s=k_2+l_2+1$.
Putting it all together, we find that $\mathcal{M}$ is: \\
\begin{equation}
    \mathcal{M} =-\bigg( \frac{1}{16 \pi^2} \bigg) \bigg( \mathcal{M}_1(r,s,m,n,k_1,k_2,l_1,l_2)\frac{z_0}{z_{12}}+\frac{1}{2}\mathcal{M}_2(r,s,m,n,k_1,k_2,l_1,l_2)\bigg) d\overline{z}_0
\end{equation} \\
where we've defined: \\
\begin{equation*}
    \mathcal{M}_1 = \underset{ac}{\tilde{\sum}} \frac{(m+n-a-c)!(a+c-1)!a!c!(r+m-a)!(s+n-c)!(1+k_1+k_2)!(1+l_1+l_2-a-c)!}{(m+n)!(1+m+r+n+s-a-c)!}
\end{equation*} \\
\begin{equation*}
    \mathcal{M}_2 = \underset{ac}{\tilde{\sum}}\frac{k_1+k_2-l_1-l_2+a+c}{2+m+r+n+s-a-c}  \mathcal{M}_1^{ac}
\end{equation*} \\
\begin{equation*}
    \underset{ac}{\tilde{\sum}} =  \frac{1}{k_1!k_2!l_1!l_2!}\underset{ac}{\sum} 
\end{equation*}
where the $\mathcal{M}^{ac}$ are the coefficients in the $\mathcal{M}_1$ sum. \\ \\ \\
Using these results, we find that the one-loop corrections have the general form \footnote{Note that the structure of the double pole in is agreement with the results found in \cite{ON_THE_ASSOCIATIVITY_OF_1_LOOP_CORRECTIONS_TO_THE_CELESTIAL_OPERATOR_PRODUCT_IN_GRAVITY}.} \\ 
\begin{equation*}
    \begin{aligned}
         J_b[r,s](z)J_a[m,n](w) &\sim \overset{k_2+l_2=n+s-1}{\overset{k_1+l_1=m+r-1}{\sum_{k_i,l_i \geq 0}}}\frac{-1}{16\pi^2 (z-w)} K^{ef}(f^c_{ae} f^d_{bf} \overset{\vec{k} \vec{l}}{\mathcal{M}_1} +f^d_{ae} f^c_{bf} \overset{\vec{l} \vec{k}}{\mathcal{M}_1}) :J_c[k_1,k_2] \tilde{J}_d[l_1,l_2]:(w)\\ 
    &-\frac{\overset{\vec{k} \vec{l}}{\mathcal{M}_2} }{8 \pi^2}  h^{\vee} f^c_{ab} \bigg( \frac{1}{(z-w)^2}+\frac{1}{2(z-w)}\partial\bigg) \tilde{J}_c[m+r-1,n+s-1](w)
    \end{aligned}
\end{equation*} \\ 
\begin{equation*}
     \tilde{J}_b[r,s](z)J_a[m,n](w) \sim \overset{k_2+l_2=n+s-1}{\overset{k_1+l_1=m+r-1}{\sum_{k_i,l_i \geq 0}}}\frac{-\overset{\vec{k} \vec{l}}{\mathcal{M}_1}}{16\pi^2 (z-w)} K^{ef}f^c_{ae} f^d_{bf} :\tilde{J}_c[k_1,k_2] \tilde{J}_d[l_1,l_2]:(w)
\end{equation*}
\newpage
\nocite{*}
\bibliographystyle{JHEP}
\bibliography{One-loop_corrections_to_the_celestial_chiral_algebra_from_Koszul_duality}
\end{document}